\documentclass[aps,prc,superscriptaddress,showpacs,floatfix,nofootinbib,notitlepage,twocolumn]{revtex4-1}
\usepackage{amsmath,graphicx,float,hyperref,xcolor}
\usepackage{physics,amssymb}

\newcommand{\avg}[1]{\left\langle#1\right\rangle}

\begin{document}

\title{Cumulants of conserved charges in hydrodynamic simulations}

\author{Renan Hirayama}
\affiliation{Helmholtz Forschungsakademie Hessen f\"ur FAIR (HFHF)\\ $^1$Frankfurt Institute for Advanced Studies, Ruth-Moufang-Strasse 1, 60438 Frankfurt am Main, Germany}
\author{Fr\'ed\'erique Grassi}
\affiliation{Instituto de F\`\i sica, Universidade de S\~ao Paulo, Rua do Mat\~ao 1371, 05508-090 S\~ao Paulo-SP, Brazil}
\author{Willian Matioli Serenone} 
\affiliation{Instituto de F\`\i sica, Universidade de S\~ao Paulo, Rua do Mat\~ao 1371, 05508-090 S\~ao Paulo-SP, Brazil}
\author{Jean-Yves Ollitrault}
\affiliation{Universit\'e Paris Saclay, CNRS, CEA, Institut de physique th\'eorique, 91191 Gif-sur-Yvette, France}

\begin{abstract}
We introduce a fast and simple method of computing cumulants of net-proton or net-charge fluctuations in event-by-event hydrodynamic simulations of heavy-ion collisions. 
One evaluates the mean numbers of particles in every hydrodynamic event. 
Cumulants are then expressed as a function of these mean numbers. 
We implement the corrections due to global conservation laws. 
The method is tested using ideal hydrodynamic simulations of Au+Au collisions at $\sqrt{s_{NN}}=200$ AGeV with the NeXSPheRIO code. 
Results are in good agreement with experimental data on net-proton and net-charge fluctuations by the STAR collaboration. 
\end{abstract}
\date{\today}

\maketitle


\section{Introduction}
One of the primary goals of relativistic heavy-ion collisions is to study experimentally the thermodynamic properties of the theory of strong interactions, quantum chromodynamics (QCD). 
In particular, one hopes to relate the event-to-event fluctuations of a physical quantity, for example the net electric charge observed in a specific detector~\cite{STAR:2014egu}, to the thermodynamic fluctuations of that quantity at some temperature. 
The interest in thermodynamic fluctuations is twofold. 
First, they diverge in the vicinity of a phase transition~\cite{Stephanov:1998dy} and might help locate a critical point in the QCD phase diagram~\cite{Fukushima:2010bq}. 
This is one of the main goals of the beam energy scan at the Relativistic Heavy Ion Collider (RHIC)~\cite{Luo:2017faz}, where hints of enhanced fluctuations have been reported~\cite{STAR:2020tga,STAR:2021iop}. 
Second, at zero baryon chemical potential, thermodynamic fluctuations can be calculated from first principles in lattice QCD~\cite{Borsanyi:2011sw,Borsanyi:2018grb}, and these calculations have been compared with experimental measurements~\cite{Bazavov:2012vg,Borsanyi:2013hza,Alba:2014eba,Bazavov:2020bjn}.

Hydrodynamic models~\cite{Gale:2013da,Romatschke:2017ejr} have been notoriously successful in reproducing bulk observables of relativistic heavy-ion collisions~\cite{Bernhard:2019bmu,Nijs:2020roc,JETSCAPE:2020mzn}. 
Therefore, hydrodynamics seems a natural framework for modeling the fluctuations which are measured experimentally:  
One describes the strongly-coupled quark-gluon matter formed in the collision as a relativistic fluid, which expands freely and cools down until it reaches the so-called freeze-out temperature, where it is converted into hadrons. 
Typical fluctuation observables depend on the temperature, not on the fluid velocity, because they involve particle numbers, not momenta~\cite{Asakawa:2015ybt}. 
This implies that in a hydrodynamic calculation, fluctuations are essentially those of a hadron gas at the freeze-out temperature~\cite{Stephanov:1999zu}. 
Now, first-principles calculations of thermodynamic fluctuations in QCD give results in agreement with a hadron gas~\cite{Borsanyi:2014ewa} in the relevant temperature range. 
This suggests that hydrodynamics is a valid framework for investigating fluctuations. 

A full hydrodynamic calculation has significant advantages over a simple thermodynamic calculation: 
First, the modification of the momentum distribution due to the collective fluid velocity is taken into account, so that the kinematic cuts inherent to particle detectors can be properly implemented~\cite{Bluhm:2020mpc}. 
Second, effects of event-to-event fluctuations, which are not taken into account in lattice calculations~\cite{Borsanyi:2013hza}, can be modeled realistically~\cite{Socolowski:2004hw,Shen:2017bsr}. 
However, hydrodynamics is a continuous description, while fluctuation observables involve discrete particle numbers~\cite{Asakawa:2015ybt}. 
Therefore, evaluating these observables within a hydrodynamic model is not straightforward. 
The goal of this article is to introduce a simple and efficient method to evaluate cumulants in event-by-event hydrodynamics. 
We illustrate its interest by comparing theoretical calculations with experimental results for net-proton~\cite{STAR:2013gus} and net-charge~\cite{STAR:2014egu} fluctuations in Au+Au collisions at $\sqrt{s_{NN}}=200$ AGeV. 

The fully-consistent way of studying fluctuations with hydrodynamics would be to include them in the hydrodynamic description itself~\cite{Calzetta:1997aj,Kapusta:2011gt,An:2019osr,Miron-Granese:2020mbf}. 
Hydrodynamic fluctuations are large near a critical point~\cite{Jiang:2015hri,Nahrgang:2018afz,Pradeep:2021opj}, and have non-trivial effects even in the absence of a critical point~\cite{Yan:2015lfa,Akamatsu:2016llw,Sakai:2020pjw,An:2020vri}. 
Here, we postulate that fluctuations in heavy-ion collisions originate mostly from the initial state, and we simply neglect hydrodynamic fluctuations.  
We carry out a standard hydrodynamic calculation, without any hadronic afterburner~\cite{Petersen:2008dd,Song:2010mg}, and neglecting the interactions in the hadronic gas at freeze-out~\cite{Vovchenko:2020kwg,Vovchenko:2021kxx}. 
We evaluate fluctuation observables at freeze-out, in a way that matches the experimental procedure as closely as possible. 

The standard observables to characterize fluctuations are cumulants of various quantities: 
Proton or antiproton multiplicity~\cite{HADES:2020wpc,STAR:2021iop}, net-proton number~\cite{ALICE:2019nbs,STAR:2020tga,STAR:2013gus,STAR:2021rls}, net electric charge~\cite{STAR:2014egu,PHENIX:2015tkx}, net-kaon number~\cite{STAR:2017tfy,Bellwied:2018tkc}, as well as mixtures of these (off-diagonal cumulants) \cite{Ding:2015fca,STAR:2019ans,Bellwied:2019pxh}. 
In hydrodynamics, hadrons are emitted independently on the freeze-out surface. 
Therefore, the only non-trivial information returned by the hydrodynamic calculation is, for a given event, the expected value of the  number of particles in a given phase-space window. 
We express the cumulants in terms of these event-by-event expected values in Sec.~\ref{s:method}. 
We take into account the correlations arising from global conservation laws~\cite{Borghini:2000cm,Bzdak:2012an,Hammelmann:2022yso}, which have proven to be crucial in describing experimental data~\cite{ALICE:2019nbs}. 
As we shall see, the advantage of our formulation is that accurate values of cumulants can be obtained with a number of hydrodynamic events which is smaller by orders of magnitude than the number of events in an actual experiment or in a transport calculation~\cite{Hammelmann:2022yso,Xu:2016qjd}.

In Sec.~\ref{s:results}, we implement this method using ideal hydrodynamic calculations of Au+Au collisions at $\sqrt{s_{NN}}=200$ AGeV. 
We evaluate, as a function of the collision centrality, the first four cumulants of several quantities: 
Number of protons, of antiprotons, net-proton number (number of protons minus number of antiprotons), net electric charge seen in a detector. 
We compare our results with data from the STAR collaboration~\cite{STAR:2013gus,STAR:2014egu}.

\section{Evaluating cumulants in event-by-event hydrodynamics}
\label{s:method}

\subsection{Definitions}
\label{s:definitions}

We first recall some useful definitions.
The observable of interest is an integer, $N$, which is measured in every collision event. 
The event-to-event fluctuations of $N$ can be characterized by its moments $\mu_n$: 
\begin{equation}
\label{moments}
\mu_n\equiv \avg{N^n},
\end{equation}
where $n$ is a positive integer, and angular brackets denote an average over an ensemble of events with the same beam energy and ions, belonging to the same centrality class. 
The whole series of moments can be obtained through the power series expansion of the moment generating function: 
\begin{equation}
\label{MGF}
\avg{e^{zN}}= \sum_{n=0}^\infty \mu_n\frac{z^n}{n!}. 
\end{equation}
The cumulants $C_n$, with $n\ge 1$, are defined by the power series expansion of the logarithm~\cite{Asakawa:2015ybt}: 
\begin{equation}
\label{CGF}
\ln\avg{e^{zN}}= \sum_{n=1}^\infty C_n\frac{z^n}{n!},
\end{equation}
Using Eqs.~(\ref{MGF}) and (\ref{CGF}), one can express the cumulants as a function of the moments. 
For the first four cumulants, which we study in detail in Sec.~\ref{s:results}, the explicit expressions are: 
\begin{eqnarray}
\label{mu_to_C}
C_1 &=& \mu_1\cr
C_2 &=& \mu_2 - \mu_1^2\cr
C_3 &=& \mu_3 - 3\mu_2\mu_1 + 2\mu_1^3\cr
C_4 &=& \mu_4 - 4\mu_3\mu_1-3\mu_2^2+12\mu_2\mu_1^2-6\mu_1^4.
\end{eqnarray}
The first cumulants $C_1$ and $C_2$ are the mean and the variance. 
The next two cumulants, $C_3$ and $C_4$, vanish if the probability distribution of $N$ is Gaussian. 
They are called skewness and kurtosis. 

\subsection{Two-step averaging}
\label{s:twostep}

We now detail the derivation of cumulants in event-by-event hydrodynamics.
The most straightforward way to mimic experiment would be to take an initial condition, solve the hydrodynamic equations, emit hadrons on the freeze-out surface by Monte Carlo sampling, and repeat the whole process many times. 
However, solving the hydrodynamic equations is time consuming, and a faster procedure can be devised, which amounts to using the same initial conditions several times, by repeating the sampling procedure at freeze out. 

Our starting point is the observation that a hydrodynamic event is not the same thing as an experimental event, because hydrodynamics does not retain the microscopic information on particles.
A single hydrodynamic simulation with given initial conditions may lead to different hadronic emissions in the end, since hadrons are emitted randomly at freeze-out. 
Therefore, a hydrodynamic event is actually an {\it ensemble\/} of events with the same initial conditions~\cite{DerradideSouza:2015kpt}.  
Then, the average over events in hydrodynamics is a two-step process.  
First an ensemble average over events with identical initial conditions, then an average over initial conditions:
\begin{equation}
  \label{twostep}
\avg{e^{zN}}=\avg{\avg{e^{zN}}_{\rm fo}}_{\rm ic}, 
\end{equation}
where the subscript ``fo'' refers to the average over the ensemble of events at freeze-out, while the subscript ``ic'' refers to the average over hydrodynamic events. 

\subsection{Average at freeze-out}
\label{s:foaverage}

We now explain how the average at freeze-out is done in practice.
We start with the simple case where $N$ denotes the multiplicity of a specific particle, say, the proton multiplicity in a certain phase-space window. 
The only information delivered by hydrodynamics for a fixed initial condition is the expected value of $N$ at freeze-out, which we denote by $\bar N$. 
As recalled above, the hydrodynamic description assumes that hadrons are emitted independently on the freeze-out surface. 
Note that this independence strictly holds only before strong decays take place. 
The decay chains induce correlations between the decay products. 
We neglect these correlations, whose effects will be studied in a forthcoming publication, and we assume that independence is still a good approximation for the stable particles which are seen experimentally. 

For independent particle emission, the probability distribution of $N$ is a Poisson distribution, which is completely specified by $\bar N$.
This simple case is studied in Appendix~\ref{s:poisson}. 
In the case of net-charge or net-baryon fluctuations, however, it is essential to take into account the global conservation law.
The net charge and the net baryon number are fixed in every event for the whole collision system.
Experiments see fluctuations of these numbers because they only detect a fraction of the particles. 

The conservation law induces correlations between outgoing particles~\cite{Borghini:2000cm,Bzdak:2012an}.
If, in a given event, there is more charge in one region of phase space, this must be compensated by less charge in the remaining phase space. 
These correlations are not taken into account in the standard hydrodynamic description, where the conservation laws are satisfied only on average. 
Sophisticated methods have been developed for  implementing conservation laws at freeze-out~\cite{Oliinychenko:2019zfk}. 
We choose a simplified approach, by assuming that  the multiplicities of protons and antiprotons are {\it both\/} fixed for the whole collision system (in the case of net-proton fluctuations), not only their difference~\cite{Bzdak:2012an,Luo:2014tga}.
We denote by $N_{\max}$ the value of $N$ for the whole collision system,
and by $\alpha$ the probability for a proton to be seen in the detector~\cite{ALICE:2019nbs}:
\begin{equation}
\label{defalpha}
  \alpha\equiv \frac{\bar N}{N_{\max}}. 
\end{equation}
The constraint $N\le N_{\max}$ is then implemented by assuming that the probability of $N$ is a binomial distribution with success probability $\alpha$, rather than a Poisson distribution. 
The moment generating function of the binomial distribution is: 
\begin{equation}
\label{genbinomial}
\avg{e^{zN}}_{\rm fo}=(1-\alpha+ \alpha e^z)^{\bar N/\alpha},
\end{equation}
where we have used Eq.~(\ref{defalpha}) to express $N_{\max}$ in terms of $\alpha$ and $\bar N$.\footnote{Note that Eq.~(\ref{genbinomial}) can be used even if $N_{\max}$ is not an integer.}
Expanding Eq.~(\ref{genbinomial}) to first order in $z$, one finds that the average value of $N$ is $\bar N$, as it should.
Only the higher-order moments, $\avg{N^n}_{\rm fo}$  with $n\ge 2$,  depend on $\alpha$.
One recovers the Poisson distribution as a limiting case when $\alpha\ll 1$, as shown in Appendix~\ref{s:poisson}. 

The generalization to the net-proton number $N_+-N_-$, where we now denote by $N_+$ and $N_-$ the numbers of protons and antiprotons, is straightforward.
One assumes that in a given hydrodynamic event, $N_+$ and $N_-$ are independent variables.
Applying Eq.~(\ref{genbinomial}) to $N_+$ and $N_-$, one obtains: 
\begin{eqnarray}
\label{gennetbinomial}
\avg{e^{z(N_+-N_-)}}_{\rm fo}
&=&\avg{e^{zN_+}}_{\rm fo}\avg{e^{-zN_-}}_{\rm fo}\cr
&=&(1-\alpha_+ + \alpha_+  e^z)^{\frac{{\bar N_+}}{\alpha_+}}
(1-\alpha_-+ \alpha_- e^{-z})^{\frac{\bar N_-}{\alpha_-}},
\end{eqnarray}
where $\alpha_+$ and $\alpha_-$ denotes the average fractions of protons and antiprotons seen in the detector, which differ in general from one another. 

\subsection{Average over initial conditions}
\label{s:icaverage}

The last step is to average over initial conditions.
This averaging is done for the moments, according to Eqs.~(\ref{MGF}) and (\ref{twostep}).
For the sake of illustration, we provide the explicit expressions for the first two moments of the proton distribution, which are obtained by inserting Eq.~(\ref{genbinomial}) into Eq.~(\ref{twostep}): 
\begin{eqnarray}
  \label{twomomentsproton}
  \mu_1&=&\avg{\bar N}_{\rm ic}\cr
  \mu_2&=&\avg{\bar N^2}_{\rm ic}+\avg{(1-\alpha)\bar N}_{\rm ic}. 
\end{eqnarray}
Note that the fraction $\alpha$ of protons falling into the acceptance window of the detector depends slightly on the initial conditions of the hydrodynamic calculation, which is the reason why we keep the factor $(1-\alpha)$ inside the average in Eq.~(\ref{twomomentsproton}). 

The cumulants are then obtained from the moments using Eqs.~(\ref{mu_to_C}).
The variance is
\begin{eqnarray}
  \label{varianceproton}
  C_2&=&\avg{\bar N^2}_{\rm ic}+\avg{(1-\alpha)\bar N}_{\rm ic}-\avg{\bar N}^2_{\rm ic}\cr 
&=&{\rm Var}(\bar N)+\avg{(1-\alpha)\bar N}_{\rm ic},
\end{eqnarray}
where, in the last line, we have introduced the variance of $\bar N$ over initial conditions, ${\rm Var}(\bar N)\equiv \avg{\bar N^2}_{\rm ic}-\avg{\bar N}_{\rm ic}^2$.
The variance $C_2$ is the sum of two positive contributions.
The first contribution corresponds to fluctuations in initial conditions.
The second contribution is the average variance of the binomial distribution, which corresponds to fluctuations at freeze-out. 

Similarly, the moments of the distribution of the net-proton number $N_+-N_-$ are obtained by inserting Eq.~(\ref{gennetbinomial}) into Eq.~(\ref{twostep}).
The variance is 
\begin{equation}
  \label{variancenetproton}
  C_2={\rm Var}(\bar N_+-\bar N_-)+\avg{(1-\alpha_+)\bar N_+}_{\rm ic}+\avg{(1-\alpha_-)\bar N_-}_{\rm ic}.
\end{equation}
Note that the contributions of fluctuations at freeze-out add up for protons and antiprotons. 

We do not write the explicit expressions of higher-order moments and cumulants because they are more cumbersome and bring little added value, since arbitrary orders can be obtained automatically by expanding the generating function. 
Note, however, that the contributions from fluctuations in initial conditions and fluctuations at freeze-out do not appear as separate terms in higher-order cumulants.
They are intertwined in a non-trivial way. 
Therefore, initial fluctuations have a non-trivial effect on higher-order cumulants, and event-by-event hydrodynamic simulations allow us to evaluate this effect quantitatively.

\section{Implementation and comparison with STAR data}
\label{s:results}

We now illustrate the method outlined in Sec.~\ref{s:method} by carrying out an explicit calculation.
We run ideal hydrodynamic simulations using the NeXSPheRIO code~\cite{Aguiar:2001ac}, which has been instrumental in describing event-by-event flow fluctuations at RHIC energies~\cite{Andrade:2006yh,Takahashi:2009na,Gardim:2012yp}. 
Note that the standard description of heavy-ion collisions~\cite{Gale:2013da} now uses viscous hydrodynamics, rather than ideal hydrodynamics.
Shear viscosity~\cite{Karpenko:2015xea} and baryon diffusion~\cite{Denicol:2018wdp}, which are included in viscous calculations, have been shown to modify the particle spectra at RHIC energies. 
However, the present study focuses on fluctuations, and it is widely thought that event-by-event fluctuations largely originate from the initial state at ultrarelativistic energies~\cite{Alver:2010gr,Luzum:2013yya}. 
Thus one expects these observables to have limited sensitivity to transport coefficients, even though this should eventually been checked through explicit calculations.

\subsection{Hydrodynamic setup}
\label{s:rescaling}

\begin{figure*}[ht]
\centering
\includegraphics[width=.45\linewidth]{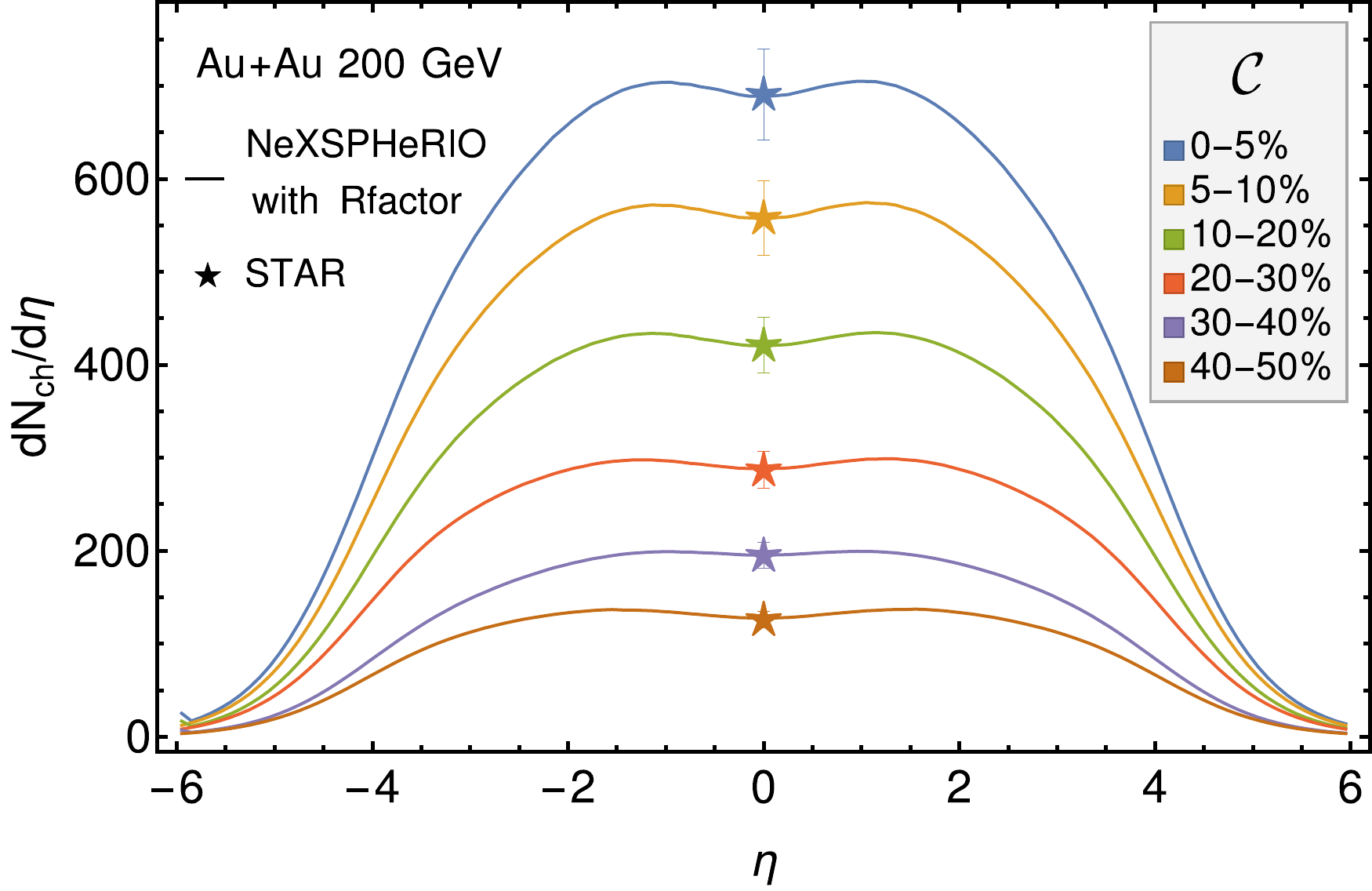}%
\hspace{0.05\linewidth}
\includegraphics[width=.45\linewidth]{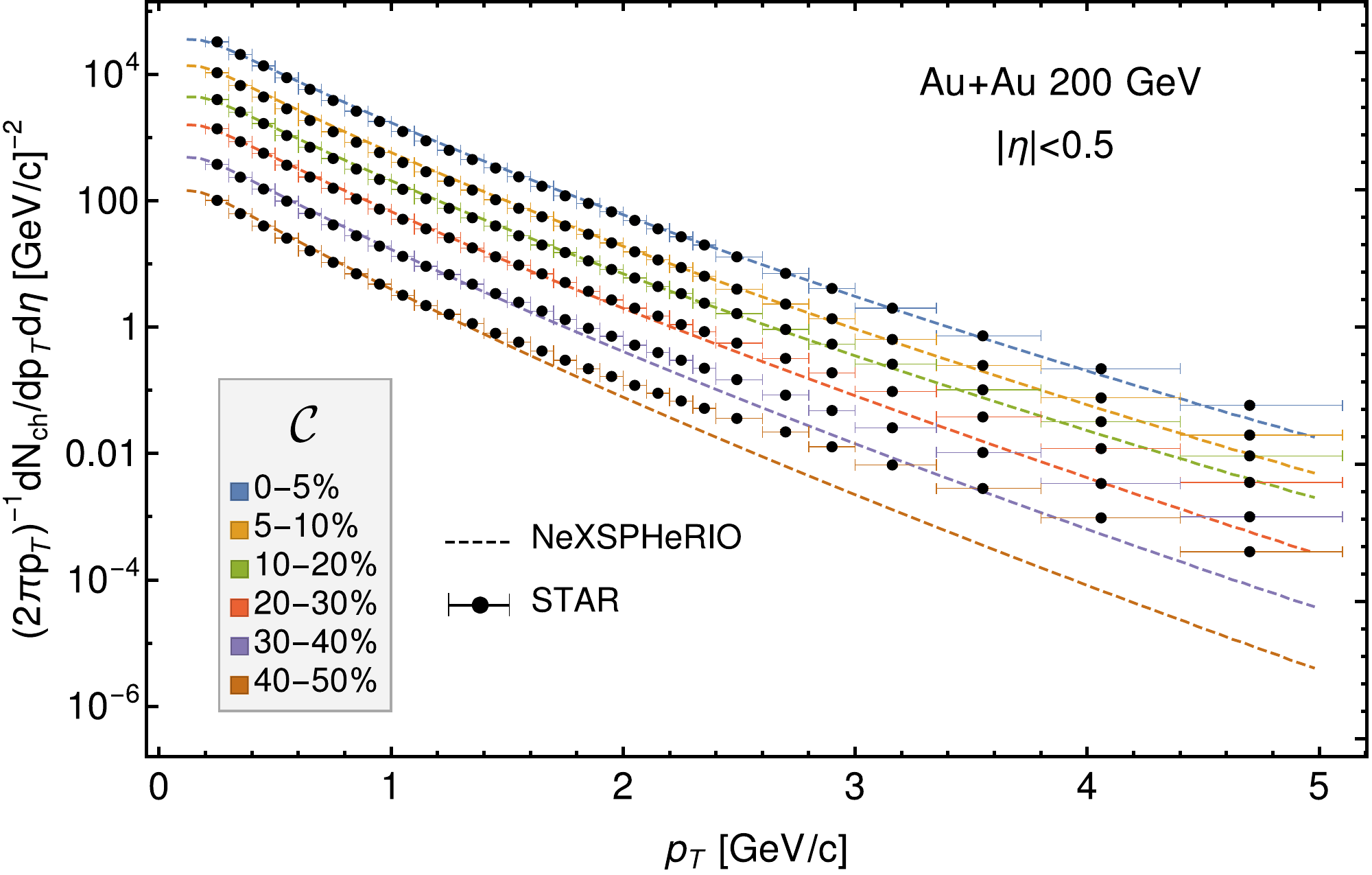}
\caption{Left: pseudorapidity distribution of charged hadrons in various centrality windows from our NeXSPheRIO calculation (lines) and experimental data from STAR~\cite{STAR:2008med}. 
Right:  transverse momentum spectra of charged hadrons in the pseudorapidity window $|\eta|<0.5$  from NeXSPheRIO  (lines) and data from STAR~\cite{STAR:2003fka}.
}
\label{fig:right_yields}
\end{figure*}	

We simulate $10^4$ Au+Au collisions at $\sqrt{s_{NN}}=200$ AGeV, with number of participants in the range $67\leq N_\text{part}\leq 394$, corresponding to the 50\% most central collisions~\cite{STAR:2008med}. 
We sort these events into 1\% centrality classes, defined according to the number of participants.
This binning is fine enough that we need not implement the ``centrality bin width correction'' (CBWC) method~\cite{Luo:2017faz}, which is used to correct for volume fluctuations when the analysis is done in wide centrality bins. 
Note that because of this fine binning, we only have 200 events per centrality bin. 
As will be clear from the results presented below, this is enough to obtain accurate results for cumulants up to $C_4$. 

In the NeXSPheRIO hydrodynamic code, the event-by-event initial conditions are borrowed from the NEXUS model~\cite{Drescher:2000ec}. 
We modify these initial conditions in the following way: 
We multiply the initial energy density with a function (Rfactor) depending on space-time rapidity, which is the same for all events in a centrality class, and which is tuned in such a way that the pseudorapidity distributions of outgoing particles has the same shape as in experiment~\cite{BRAHMS:2001llo}. 
We then run ideal hydrodynamics with this modified initial condition. 
Finally, the freeze-out temperature is adjusted in such a way that transverse momentum spectrum of charged hadrons matches experimental data in the soft sector. 
The resulting freeze-out temperature increases mildly with the centrality percentile, from 128~MeV to 142~MeV. 
The pseudorapidity and transverse momentum distributions of charged hadrons are displayed in Fig.~\ref{fig:right_yields}. 
We only show experimental data from the STAR collaboration for the sake of consistency with the following results shown in this section, where we show results on fluctuations which are also from STAR. 

The emission of hadrons on the freeze-out surface is done via Monte Carlo. 
As explained in Sec.~\ref{s:foaverage}, we need to evaluate the expected value of each of the relevant multiplicities, which we have denoted by $\bar N$, in every hydrodynamic event. 
In order to reach the desired accuracy on $\bar N$,  we repeat the Monte Carlo sampling~\cite{Gardim:2011qn} 2500 times for each hydrodynamic event.

\subsection{Cumulants of proton, antiproton and net-proton numbers}
\label{s:protoncumulants}

The STAR Collaboration has published data on the first four cumulants ($C_1$ to $C_4$) for protons, antiprotons, and the net-proton number, as a function of the collision centrality~\cite{STAR:2013gus}. 
In order to evaluate these cumulants in NeXSPheRIO, as explained in Sec.~\ref{s:method}, we need to evaluate the mean numbers of protons and antiprotons $\bar N_+$ and $\bar N_-$ in each hydrodynamic event, with the same kinematic cuts as in experiment. 
We also need to evaluate the corresponding acceptance fractions $\alpha_+$ and $\alpha_-$, defined by Eq.~(\ref{defalpha}). 
We explain how these quantities are obtained.  

\begin{figure}[ht]
\centering
\includegraphics[width=.9\linewidth]{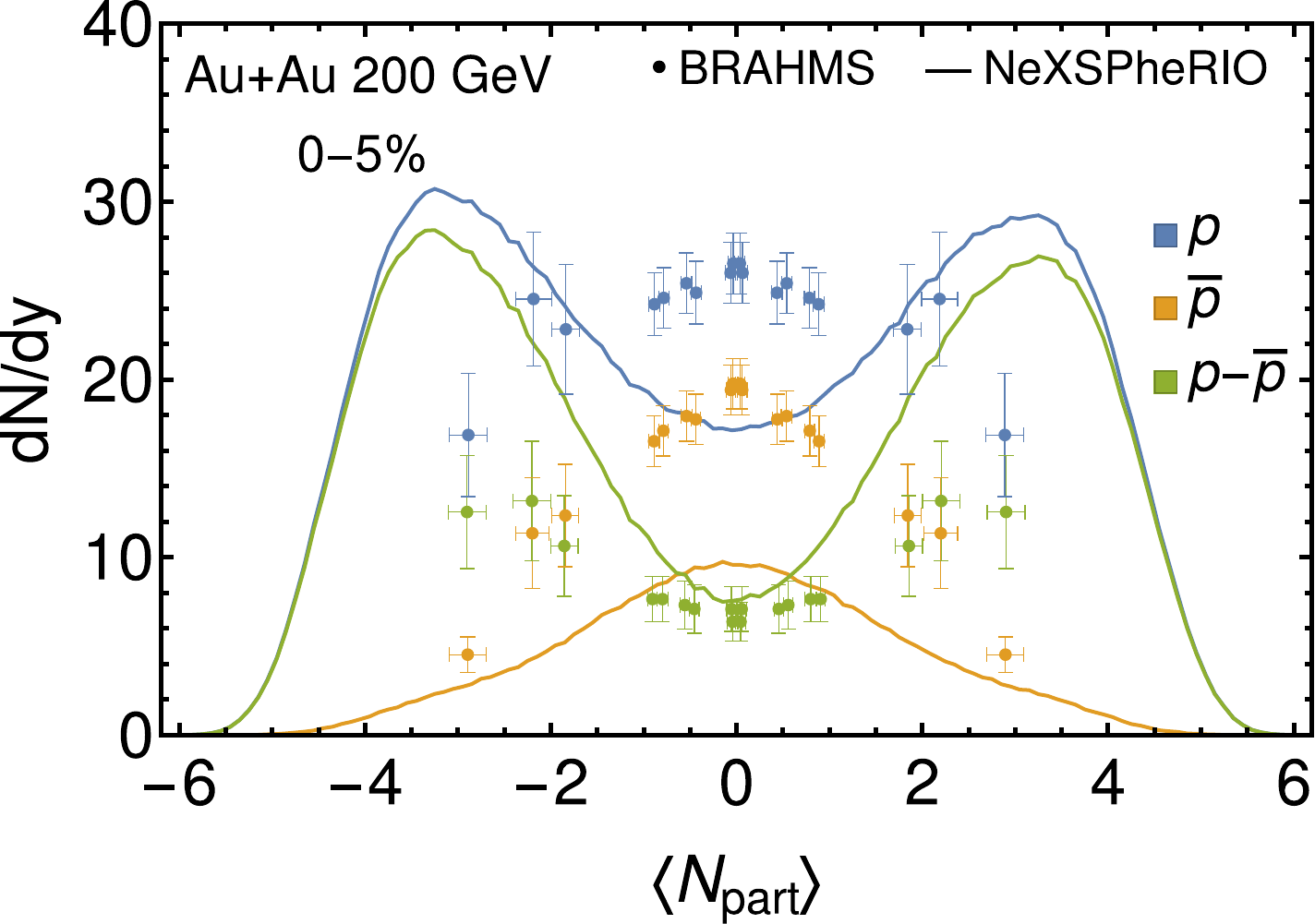}%

\includegraphics[width=.9\linewidth]{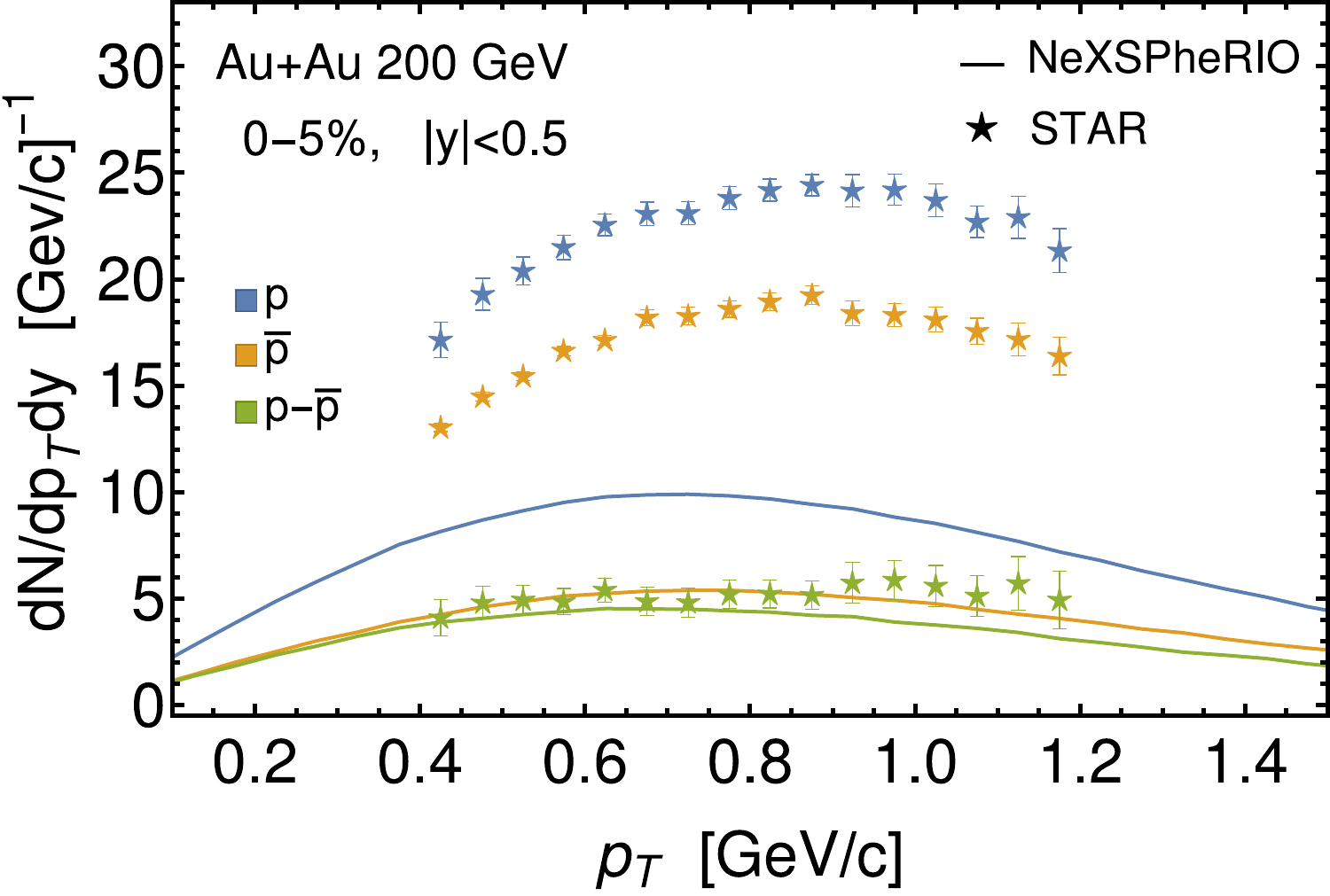}
\caption{Top: Rapidity distribution of protons, antiprotons and net-proton number in the 0-5\% centrality window. 
Lines: NeXSPheRIO results. Symbols: experimental data from BRAHMS~\cite{BRAHMS:2001llo}. 
Bottom: Transverse momentum distribution of protons, antiprotons, and net-proton number near midrapidity (note: vertical axis is not logarithmic). 
Lines: NeXSPheRIO results.  Symbols: data from STAR \cite{STAR:2008med}. 
}\label{fig:wrong_yields}
\end{figure}	

\begin{figure*}[ht]
\centering
\includegraphics[width=.75\linewidth]{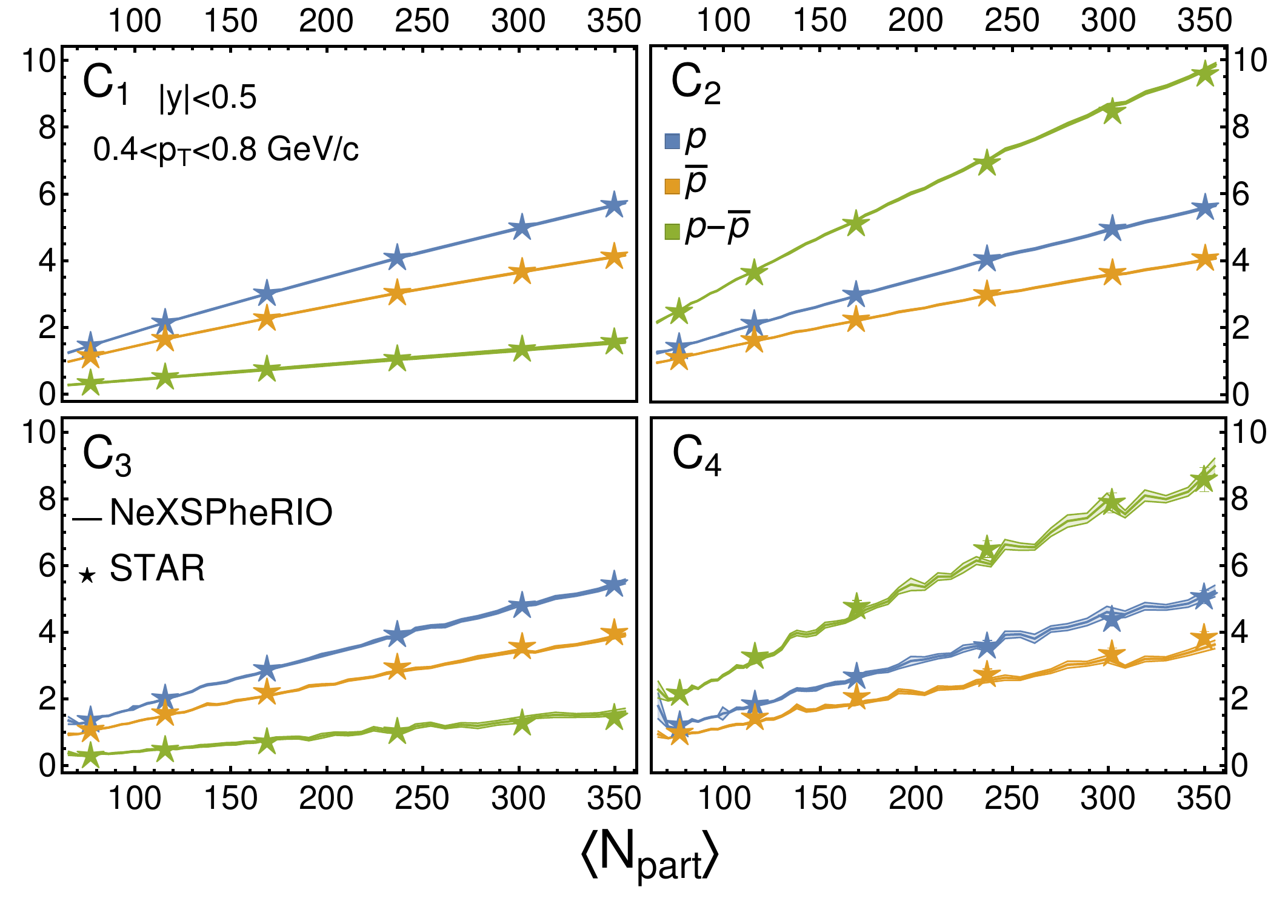}
\caption{(Color online) Cumulants ($C_1$ to $C_4$) of $p, \bar{p},$ and $p-\bar{p}$ in Au+Au collisions at $\sqrt{s_{NN}}=200$ GeV from our calculation (solid lines) and experiment (symbols)~\cite{STAR:2013gus}, as a function of the number of participants which measures the collision centrality. 
Note that the calculation has been rescaled so as to match experiment for $C_1$ (see text for details).
Error bands in simulation are evaluated by jackknife resampling. 
}\label{fig:Cn_p_rescaled}
\end{figure*}

We cannot borrow the mean values of proton and antiproton multiplicities directly from our hydrodynamic calculation, because the calculation is in  agreement with experiment for the net-proton number but not for protons and antiprotons individually. 
The reason is that NeXSPheRIO was tuned to reproduce the net-proton yield near mid-rapidity.  
This is illustrated in Fig.~\ref{fig:wrong_yields}. 
The left panel shows that our model of initial conditions underestimates baryon stopping, resulting in a proton spectrum which is too wide. 
In addition, the number of antiprotons is underpredicted. 
This is due to the fact that our freeze-out temperatures are lower than the temperature of chemical equilibration~\cite{Andronic:2017pug}, at which relative abundances are reproduced. 
Since we do not implement partial chemical equilibrium~\cite{Huovinen:2007xh,Huovinen:2009yb}, the number of heavy hadrons, such as antiprotons, is too small.  

In order for the comparison with data to be meaningful, we rescale the calculated $\bar N_+$ and $\bar N_-$.
Specifically, we multiply the calculated values of $\bar N_+$ and $\bar N_-$ by a constant which we adjust in such a way that the average values over events, denoted by $C_1$, coincide with the experimental values in each centrality window. 
Since the centrality binning is finer in our calculation than in experiment, we use linear interpolation to evaluate these multiplicative constants between two data points. 

We then evaluate the fractions of protons and antiprotons, $\alpha_+$ and $\alpha_-$, falling into the acceptance window chosen by the STAR analysis, specifically, the rapidity window $|y|<0.5$ and the transverse momentum window $0.4<p_T<0.8$~GeV/c. 
We proceed as follows. 
We first interpolate the rapidity distribution of protons and antiprotons from experiment (left of Fig.~\ref{fig:wrong_yields}) using the three-source model of Ref.~\cite{Gao:2016czp}. 
We then evaluate the fraction in  $|y|<0.5$, which is $\alpha_{y,+}=0.153$ for protons, and  $\alpha_{y,-}=0.203$ for antiprotons. 
We assume that these fractions are independent of centrality. 
Next, we fit the $p_T$ spectra (right of Fig. \ref{fig:wrong_yields}) with blast-wave fits \cite{Schnedermann:1993ws}. 
We then evaluate the fraction in the window $0.4<p_T<0.8$~GeV/c, which is $\alpha_{p_T}= 0.275$ for both protons and antiprotons. 
Particles are detected if both $y$ and $p_T$ satisfy cuts, so that one must multiply the corresponding fractions $\alpha_{y,\pm}$ and $\alpha_{p_T}$. 
We eventually obtain $\alpha_+=0.042$ for protons, and $\alpha_-=0.056$ for antiprotons.\footnote{Our calculation actually uses a value of $\alpha_\pm$ which fluctuates event to event, and is evaluated in the following way. 
We evaluate $\alpha_\pm$ using Eq.~(\ref{defalpha}) for each hydrodynamic event. 
We then rescale it in such a way that the average over events matches the value calculated from measured spectra. 
We have checked that our results are essentially insensitive to these event-to-event fluctuations of $\alpha_\pm$.}

Our results are presented in Fig.~\ref{fig:Cn_p_rescaled}. 
We emphasize that the first panel, corresponding to $C_1$, is an input of our calculation, as explained above. 
The non-trivial output is represented by the higher-order cumulants $C_2$, $C_3$ and $C_4$.
One first notes that the statistical error on our results, represented as a band, is very small, despite the small number of hydrodynamic events.
This illustrates that our method is an efficient way of evaluating cumulants in hydrodynamics. 

\begin{figure*}[ht]
\centering
\includegraphics[width=.8\linewidth]{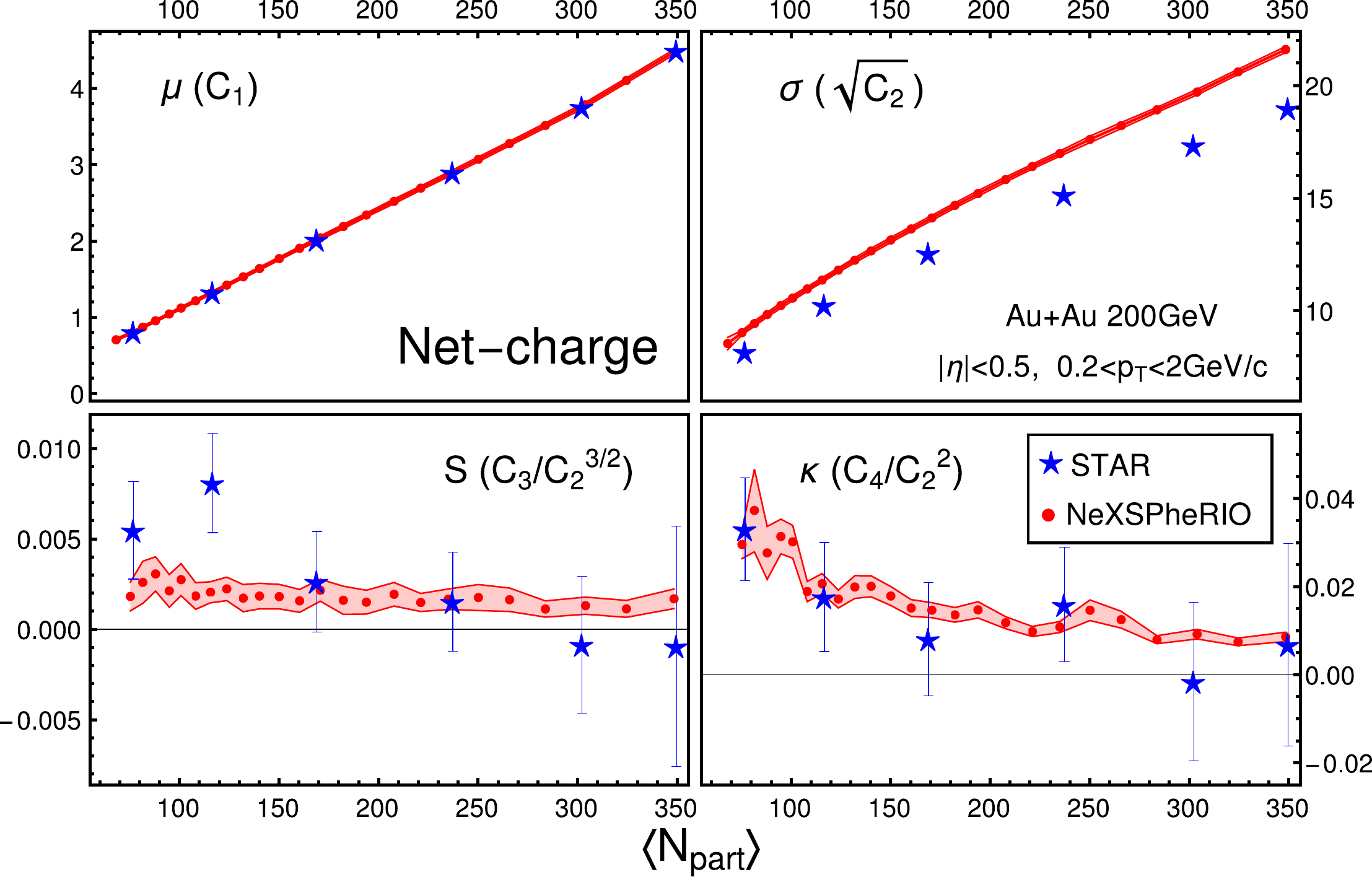}
\caption{(Color online) Centrality dependence of cumulants of the net charge in Au+Au collisions at $\sqrt{s_{NN}}=200$ GeV, from our hydrodynamic calculation (solid lines) and experiment (symbols)~\cite{STAR:2014egu}.
As in Fig.~\ref{fig:Cn_p_rescaled}, the calculated $C_1$ has been rescaled to match data (see text for details). 
} \label{fig:stats_netchg_rescaled}
\end{figure*}

Agreement with experiment is excellent. 
Higher-order cumulants are dominated by Poisson fluctuations, as detailed in Appendix~\ref{s:poisson}. 
The corrections to this baseline~\cite{Luo:2014tga} come on the one hand from the global conservation law, and on the other hand from initial-state fluctuations. 
The global conservation law, quantified by the parameters $\alpha_+$ and $\alpha_-$, decreases the cumulants, as illustrated by Eqs.~(\ref{varianceproton}) and (\ref{variancenetproton}) for the variance.  
While the correction is only $\sim 5\%$ for $C_2$, it is larger for higher order cumulants, becoming as large as $30\%$ for $C_4$.\footnote{Specifically, the cumulants of the binomial distribution, which are obtained by inserting Eq.~(\ref{genbinomial}) into Eq.~(\ref{CGF}), are smaller than those of the Poisson distribution by a factor $(1-\alpha)$ for $C_2$, $(1-\alpha)(1-2\alpha)$ for $C_3$, $(1-\alpha)(1-6\alpha+6\alpha^2)$ for $C_4$.} 
Initial-state fluctuations, on the other hand, {\it increase\/} the cumulants. 
The effects of initial-state fluctuations and global conservation almost cancel each other out, so that the cumulants are eventually very similar to the expectation from Poisson fluctuations. 
This seems to be a numerical coincidence. 
The conclusion is that the almost perfect agreement between our calculation and experimental results is less trivial than it looks. 
This will be illustrated further in Appendix~\ref{s:poisson}. 

\subsection{Cumulants of net charge}
\label{s:netchargecumulants}

The STAR collaboration has also evaluated the cumulants of the net electric charge~\cite{STAR:2014egu}, defined as the difference between positive and negative charged particle multiplicities, within the pseudorapidity window $|\eta|<0.5$ and transverse momentum window $0.2 < p_T < 2.0$~GeV/c. 
The evaluation of these cumulants in NeXSPheRIO is again carried out along the lines of Sec.~\ref{s:method}, where $N_+$ and $N_-$ now denote the numbers of positive and negative charge particles seen in the event. 
We need to evaluate their mean values $\bar N_+$ and $\bar N_-$ in each hydrodynamic event, with the same kinematic cuts as in experiment, and the corresponding acceptance fractions $\alpha_+$ and $\alpha_-$. 

As shown in Fig.~\ref{fig:right_yields}, our code correctly reproduces the distributions of all charged particles, therefore we take the sum  $\bar N_++\bar N_-$ directly from our calculation. 
However, for the same reason as our code does not reproduce the yields of identified particles, it underestimates (by $\sim 25\%$) the mean net charge at central rapidity. 
Therefore, we multiply the difference $\bar N_+-\bar N_-$ from our calculation by a constant which we adjust in such a way that the average value over events, denoted by $C_1$, coincides with the experimental values in every centrality window. 

We then evaluate the fraction $\alpha$ of charged particles within the kinematic cuts. 
We assume for simplicity that this fraction is identical for negatively and positively charged particles, that is, $\alpha_+\simeq\alpha_-$, and we estimate it using the measured pseudorapidity distribution of charged particles~\cite{BRAHMS:2001llo}. 
The resulting value is $\alpha\simeq 0.09$ and varies mildly with centrality.\footnote{As the collision becomes more central, $\alpha$ increases slightly due to increased stopping. We obtain $\alpha=0.0975$, $0.0944$, $0.0924$, $0.0901$, $0.0884$, $0.0858$ in the 0-5, 5-10, 10-20, 20-30, 30-40, 40-50\% centrality windows.}

The width of centrality bins in our calculation is 2\% for this analysis, instead of 1\% in Sec.~\ref{s:protoncumulants}. 
This increases the number of hydrodynamic events per bin and reduces the statistical error. 
It is somewhat counter-intuitive that one needs more statistics for an analysis which uses many more particles (all charged particles as opposed to just baryons, and in a much wider transverse momentum window). 
This is one of the paradoxes of cumulant analyses.

Our results are presented in Fig.~\ref{fig:stats_netchg_rescaled}, together with experimental data from the STAR Collaboration~\cite{STAR:2014egu}.
As in Fig.~\ref{fig:Cn_p_rescaled}, the first cumulant $C_1$ is an input of our calculation, and the non-trivial output is represented by the higher-order cumulants. 
The presentation differs from  Fig.~\ref{fig:Cn_p_rescaled}, where we had plotted the cumulants $C_n$ themselves, in order to illustrate that they are dominated by Poisson fluctuations.  
In Fig.~\ref{fig:stats_netchg_rescaled}, we plot $\sigma\equiv \sqrt{C_2}$ instead of $C_2$. Physically, $\sigma$ represents the standard deviation of the net charge event to event. 
It is interesting to note that it is larger than the mean $C_1$, which illustrates the small charge asymmetry at this energy. 
Our calculation overestimates the value of $\sigma$, and we have not yet been able to spot the origin of this discrepancy. 

The higher order cumulants $C_3$ and $C_4$ are normalized by appropriate powers of $\sigma$. 
The resulting values of skewness and kurtosis, referred to as ``standardized'', measure the relative deviation to a Gaussian distribution. 
Fig.~\ref{fig:Cn_p_rescaled} shows that calculated values are in good agreement with experiment. 
The small values of the skewness and kurtosis, at the percent level, illustrate that net-charge fluctuations are almost Gaussian. 
The standardized kurtosis decreases as a function of $N_{\rm part}$. 
This is an illustration of the central limit theorem, from which one expects that fluctuations come closer to a Gaussian as the system size increases. 
If one neglects the effect of the global charge conservation, that is, if one sets the parameter $\alpha$ to zero, the kurtosis $C_4$ is larger by a factor $\sim 2$, and agreement with data is lost. 
This illustrates that the agreement of our calculation with data is a non-trivial achievement. 

\section{Discussion}
\label{s:conclusions}

We have introduced an efficient method for evaluating cumulants of conserved charges in event-by-event hydrodynamic simulations of heavy-ion collisions.
As far as we know, this is the first time that cumulants have been computed with event-by-event hydrodynamics incorporating both fluctuations in the initial conditions and in the freeze out. Both effects are numerically of the same size. 
We have illustrated the power of this method with ideal hydrodynamic simulations of Au+Au collisions at the top RHIC energy. 
Accurate results have been obtained for the first four cumulants of various quantities (proton and antiproton multiplicities, net-proton number, net electric charge)  with only a few hundred hydrodynamic events per centrality bin. 
This is smaller by orders of magnitude than the number of events in actual experiments, or in transport calculations~\cite{Hammelmann:2022yso,Xu:2016qjd,He:2017zpg}, but large enough to capture the non-trivial event-to-event fluctuations from the initial state. 
Generalization of our method to other quantities, such as the net-kaon number or off-diagonal cumulants~\cite{STAR:2019ans}, and to higher-order cumulants~\cite{STAR:2021rls} is straightforward. 

Due to discrepancies between our calculation and experiment, concerning the spectra of identified particles, we had to rescale the average number of particles in the acceptance window, corresponding to the first cumulant $C_1$, so as to match data. 
With this rescaling, calculated values of higher-order cumulants ($C_2$ to $C_4$) are in excellent agreement with experiment for net-proton fluctuations, and in fair agreement for net-charge fluctuations. 
The is a non-trivial achievement, which results from taking into account both the global conservation law, and event-to-event fluctuations. 
We are working on improving our description in such a way that no rescaling is needed. 
Specifically, we are developing an improved description based on smearing partons from the AMPT transport model, inspired on what was done in~\cite{Chattopadhyay:2017bjs,Fu:2020oxj}. 
The aim is  to correctly describe both the (pseudo)rapidity distributions and transverse momentum spectra for a range of energies for charged particles,  protons  and antiprotons.

One should keep in mind that the measured cumulants are dominated by Poisson fluctuations, which are essentially trivial. 
Poisson fluctuations can be removed by replacing  moments with factorial moments~\cite{Kitazawa:2017ljq}. 
This is achieved by replacing $e^z$ with $1+z$ in the left-hand side of Eqs.~(\ref{MGF}). 
$\mu_n$ in the right-hand side of Eqs.~(\ref{MGF}) then becomes the factorial moment, which is the average number of $n$-tuples per event. 
Factorial cumulants are related to factorial moments in the same way as cumulants are related to moments, i.e., through Eq.~(\ref{mu_to_C}). 
The factorial cumulant of order $n$ measures the $n$-particle correlation~\cite{Bzdak:2016sxg,DiFrancesco:2016srj},  and can be used for probing critical  fluctuations~\cite{Ling:2015yau}. 
Factorial cumulants can be evaluated in event-by-event hydrodynamics along the same line as cumulants.  
One need simply replace  $e^z$ with $1+z$ in Eqs.~(\ref{twostep}) and (\ref{genbinomial}). 
Note, finally, that the definition of factorial cumulants can be amended to remove not only Poisson fluctuations, but also the trivial correlation from the global conservation law~\cite{Rogly:2018kus}, so as to isolate non-trivial correlations and effects of initial state fluctuations. 
We intend to use these modified cumulants in conjunction with an equation of state with an adjustable critical point position~\cite{Parotto:2018pwx,Karthein:2021nxe}. 

\section*{Acknowledgments} 
This work is supported by Funda\c{c}\~ao de Amparo \`a Pesquisa do Estado de S\~ao Paulo (FAPESP) through  grants  2018/24720-6, 2018/14479-0, 2021/01670-6 and project INCT-FNA Proc.~No.~464898/2014-5, and by the Helmholtz Research Academy Hesse for FAIR (HFHF).

\appendix

\section{Poisson distribution}
\label{s:poisson}

In this Appendix, we study the simple case where particles are emitted independently at freeze-out, neglecting the correlation from the global conservation law. 
Independent particle emission implies that the distribution of any multiplicity $N$ at freeze-out is a Poisson distribution.
The moment generating function is: 
\begin{equation}
\label{genpoisson}
\avg{e^{zN}}_{\rm fo}=\exp\left((e^z-1)\bar N\right),
\end{equation}
where $\bar N$ is the mean value of $N$.
An elementary calculation shows that Eq.~(\ref{genbinomial}) reduces to Eq.~(\ref{genpoisson}) for $\alpha\ll 1$. 
This shows that the binomial distribution reduces to a Poisson distribution in this limit.

If there are no fluctuations in initial conditions, then, inserting Eq.~(\ref{genpoisson}) into Eq.~(\ref{CGF}), one obtains immediately the cumulants of the distribution of $N$ : 
\begin{equation}
\label{cumulantspoisson}
C_n=\bar N.
\end{equation}
The cumulants of the Poisson distribution are all equal. 
This is approximately true for the measured cumulants of the numbers of protons and antiprotons in Fig.~\ref{fig:Cn_p_rescaled} (circles and squares), where one readily notices that $C_1\approx C_2\approx C_3\approx C_4$ for all centralities. 
This suggests that Poisson fluctuations are the dominant source of fluctuations. 

If the multiplicities of protons and antiprotons, $N_+$ and $N_-$, are independent and both follow a Poisson distribution, then the moment generating function of the net-proton number $N_+-N_-$ is: 
\begin{equation}
\label{gennetpoisson}
\avg{e^{z(N_+-N_-)}}_{\rm fo}=\exp\left((e^z-1)\bar N_++(e^{-z}-1)\bar N_-\right). 
\end{equation}
If there are no fluctuations in initial conditions, then, inserting Eq.~(\ref{gennetpoisson}) into Eq.~(\ref{CGF}), one obtains the cumulants of the distribution of $N_+-N_-$ (known as the Skellam distribution): 
\begin{eqnarray}
\label{cumulantsskellam}
C_{2n-1}&=&\bar N_+-\bar N_-\cr
C_{2n}&=&\bar N_++\bar N_-. 
\end{eqnarray}
Looking at the measured cumulants of the net-proton number in Fig.~\ref{fig:Cn_p_rescaled} (stars), one notices that $C_1\approx C_3$ and $C_2\approx C_4$, which again suggests that Skellam fluctuations dominate. 

\begin{figure*}[ht]
\centering
\includegraphics[width=.75\linewidth]{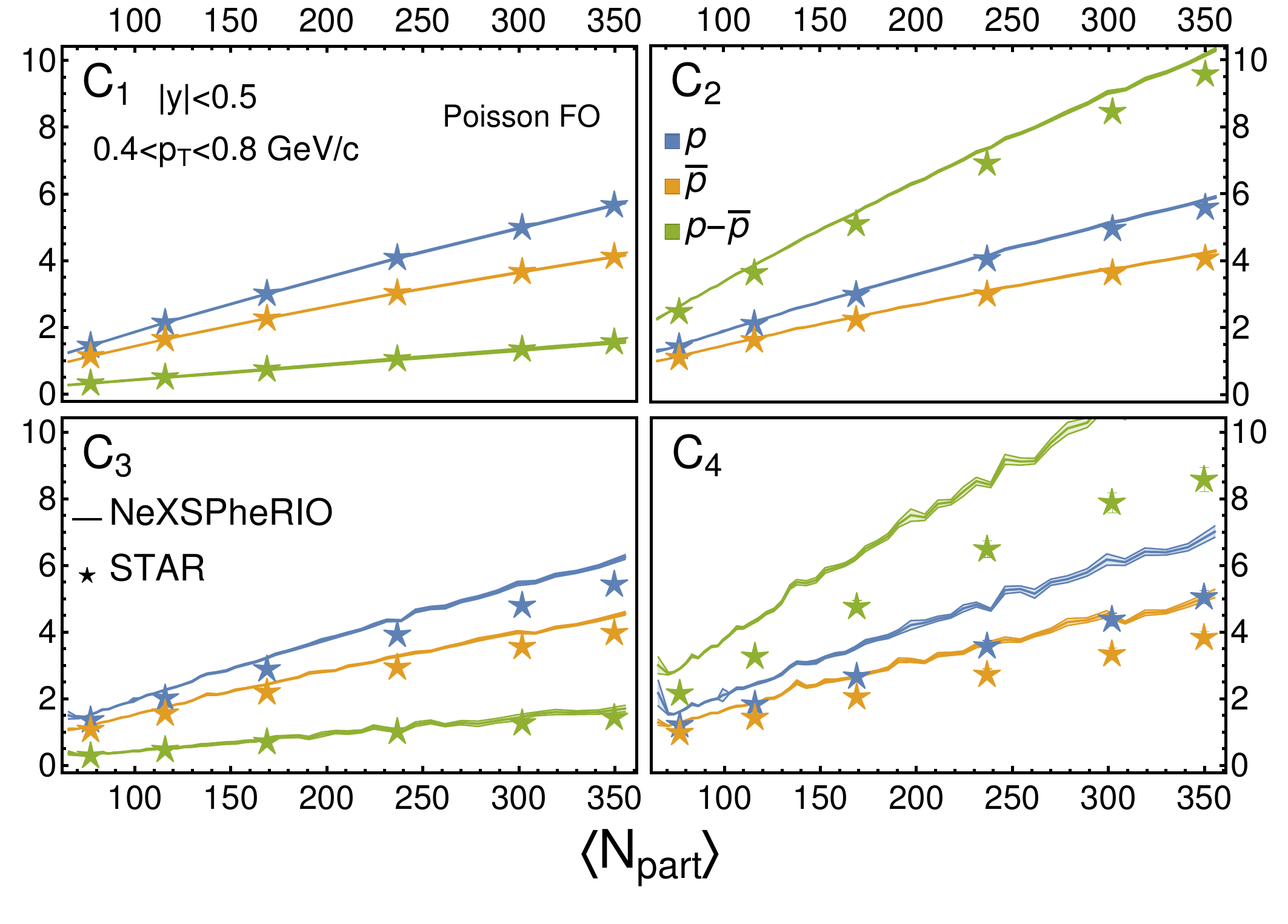}
\caption{(Color online) 
Same as Fig.~\ref{fig:Cn_p_rescaled}, where the lines now denote the results of our calculations assuming independent particle emission at freeze out, i.e., neglecting the correlations from the global conservation law.
}\label{fig:Cn_p_rescaled_poisson}
\end{figure*}
In event-by-event hydrodynamics, however, initial-state fluctuations have a sizable effect, and break the simple equalities  (\ref{cumulantspoisson}) and (\ref{cumulantsskellam}). 
This is illustrated by Fig.~\ref{fig:Cn_p_rescaled_poisson}, in which the lines display the results of our event-by-event hydrodynamic calculation, assuming independent particle emission at freeze-out. 
Fluctuations in initial conditions increase higher-order cumulants, in particular $C_4$, so that our calculation overshoots the data. 
Agreement with data is only restored after taking into account the correlations from the global conservation law, as shown in  Fig.~\ref{fig:Cn_p_rescaled}.

\end{document}